\documentclass[fleqn,usenatbib]{mnras}

\usepackage{newtxtext,newtxmath}
\usepackage[T1]{fontenc}
\usepackage{ae,aecompl}

\usepackage{graphicx}	% Including figure files
\usepackage[dvipsnames]{xcolor}	% Including figure files
\usepackage{amsmath}	% Advanced maths commands
\usepackage{amssymb}	% Extra maths symbols

\usepackage{savesym}
\savesymbol{tablenum}
\usepackage{siunitx}
\restoresymbol{SIX}{tablenum}

\newcommand{\ms}{M$_\odot$}
\newcommand{\mj}{M$_{\rm J}$}
\newcommand{\map}{$M_{\rm p}$}
\newcommand{\s}{s$^{-1}$}
\newcommand{\yr}{yr$^{-1}$}
\newcommand{\md}{$\dot{M}_{\rm wind}$}

\newcommand{\sd}{$\dot{\Sigma}_{\rm wind}$}

\title[Disc photoevaporation and gas giant migration]{The comparative effect of FUV, EUV and X-ray disc photoevaporation on gas giant separations}

\author[Jennings, Ercolano \& Rosotti]{
Jeff Jennings$^{1*}$,
Barbara Ercolano$^{1,2}$
and Giovanni P. Rosotti$^{3}$
\\
$^{1}$University Observatory Munich, Scheinerstr.1, D-81679 Munich, Germany\\
$^{2}$Excellence Cluster Origin and Structure of the Universe, Boltzmannstr.2, 85748 Garching, Germany\\
$^{3}$Institute of Astronomy, University of Cambridge, Madingley Road, Cambridge, UK\\
$^{*}${\rm jennings@usm.lmu.de}
}

\date{Accepted XXX. Received YYY; in original form ZZZ}

\pubyear{2018}

\begin{document}
\label{firstpage}
\pagerange{\pageref{firstpage}--\pageref{lastpage}}
\maketitle

\begin{abstract}
Gas giants' early ($\lesssim 5$ Myr) orbital evolution occurs in a disc losing mass in part to photoevaporation driven by high energy irradiance from the host star. This process may ultimately overcome viscous accretion to disperse the disc and halt migrating giants by starving their orbits of gas, imprinting on giant planet separations in evolved systems. Inversion of this distribution could then give insight into whether stellar FUV, EUV or X-ray flux dominates photoevaporation, constraining planet formation and disc evolution models.
We use a 1D hydrodynamic code in population syntheses for gas giants undergoing Type II migration in a viscously evolving disc subject to either a primarily FUV, EUV or X-ray flux from a pre-solar T Tauri star. 
The photoevaporative mass loss profile's unique peak location and width in each energetic regime produces characteristic features in the distribution of giant separations: a severe dearth of $\lesssim$ 2 {\mj} planets interior to 5 AU in the FUV scenario, a sharp concentration of $\lesssim$ 3 {\mj} planets between $\approx 1.5 - 2$ AU in the EUV case, and a relative abundance of $\approx 2 - 3.5$ {\mj} giants interior to 0.5 AU in the X-ray model. 
These features do not resemble the observational sample of gas giants with mass constraints, though our results do show some weaker qualitative similarities.
We thus assess how the differing photoevaporative profiles interact with migrating giants and address the effects of large model uncertainties as a step to better connect disc models with trends in the exoplanet population.
\end{abstract}


\section{Introduction}
A newly formed giant planet migrates inward due to interaction with the gaseous disc in which it is embedded \citep[e.g.,][]{Goldreich1980,Lin1986} until the gas disc dissipates. The planet's separation at the time of disc dispersal sets the initial conditions for any subsequent dynamic evolution \citep{Davies2013}, and thus the mechanisms driving disc evolution and dispersal may leave a signature in the observed distribution of giant planet separations \citep{Alexander2012a}. A comparison of exoplanet observations with planet formation models that include dissipation of the disc also places constraints on the disc dispersal process \citep[for reviews see][]{Alexander2014,ErcolanoPascucci2017}, in turn informing planet formation and migration models \citep[e.g.,][]{Ida2004a,Bitsch2015,Mordasini2015}.

For a T Tauri star embedded in a gas and dust disc, accretion of material onto the stellar surface and chromospheric activity drive UV (FUV, 6 eV $< h \nu <$ 13.6 eV; EUV, 13.6 eV $< h \nu <$ 0.1 keV) and X-ray (0.1 keV $< h \nu <$ 2 keV) emission that heat the disc (near-)surface to launch a sub-sonic, thermal wind. The effect is concentrated at different radii in the disc in the FUV, EUV and X-ray regimes \citep[e.g.,][]{Alexander2006b,Gorti2009,Ercolano2009}. The radial profile of a purely EUV driven wind peaks sharply at the disc's gravitational radius, $r_{\rm g} = GM_*/c_{\rm s}^2 \approx 1$ AU for a 1 {\ms} star, where $c_{\rm s}$ is the isothermal sound speed \citep{Font2004}. In contrast a wind profile dominated by X-rays (with a secondary EUV component) peaks nearer to 2 AU and exhibits a broad influence out to tens of AU \citep{Owen2010}. An FUV dominated profile (with EUV and X-ray components) shows high mass loss rates between a few -- 10 AU and has a secondary peak beyond 50 AU \citep{Gorti2009a}. 

The relative strengths of these winds also differ, potentially separated by 2 orders of magnitude. For a 1 {\ms} star with constant EUV luminosity $\sim 10^{42}$ photons {\s} \citep[e.g.,][]{Pascucci2009}, 
hydrodynamic models suggest a corresponding constant mass loss rate integrated across the gas disc of {\md} $\sim 10^{-10}$ {\ms} {\yr} \citep{Font2004}. This luminosity may be overestimated by an order of magnitude, equivalently a factor of $\sqrt[]{10}$ in {\md} \citep{Pascucci2014}. X-ray luminosities of young stars are observed to lie primarily between $10^{29} - 10^{31}$ erg {\s} \citep{Preibisch2005}, yielding a median mass loss rate of {\md} $\sim 10^{-8}$ {\ms} {\yr}. FUV luminosities are thought to be time-dependent, with the predominant contribution from time-varying accretion onto the star and an integrated mass loss rate initially near {\md} $\sim 10^{-8}$ {\ms} {\yr} that degrades by 2 orders of magnitude over a gas disc lifetime of a few Myr \citep[e.g.,][]{Gorti2009}. The regimes' differing radial profiles and luminosities should result in different times and locations at which FUV, EUV and X-ray dominated photoevaporative winds significantly deplete the gas disc surface density to initiate disc dispersal. 

Models of these winds predict a removal of gas from the disc surface that excavates a \lq{}gap\rq{} (annulus) once the accretion rate onto the star \citep[decaying with time and probably starting near $\sim 10^{-6}$ {\ms} {\yr} for a $\approx 1$ {\ms} star;][]{Hartmann1998,Sicilia-Aguilar2010} reaches a factor of a few smaller than the local mass loss rate due to photoevaporation \citep{Clarke2001}. The gas interior to this gap rapidly drains onto the star on its viscous timescale, $t_\nu = r^2 / \nu \sim 10(100)$ kyr at 0.1(1.0) AU, where $\nu$ is the kinematic viscosity. This results in a \lq{}hole\rq{}, an absence of gas between the star and the inner edge of the surviving outer disc, which is now directly exposed to the stellar flux. Photoevaporation models typically alter the radial mass loss profile at this point to quickly ($\sim 100$ kyr) disperse the extant disc radially outward on a timescale commensurate with observed lifetimes of roughly a few Myr \citep[e.g.,][]{Fedele2010}.

Using the disc and planetary migration model described in Section~\ref{sec:mod}, we assess in Sections~\ref{sec:sample_cases} and~\ref{sec:charac} how photoevaporation in the pure EUV, X-ray dominated and FUV dominated regimes affects gas giant migration to identify characteristic features of each regime in the distribution of giant planet separations. Comparing qualitatively against trends in current observations, we physically motivate the differences that arise from the unique photoevaporative profiles in an approach that builds on the results of \citet{Alexander2012a} for the EUV case and \citet{Ercolano2015} for the EUV and X-ray cases to now include the FUV regime. Though note \textit{we limit our analysis to the effects of the photoevaporative profiles rather than self-consistent disc models for each of the three regimes.}
Characterizing the sensitivity of our results to model uncertainties and limitations in Section~\ref{sec:unc}, we summarize our findings in Section~\ref{sec:conc} in the context of future theoretical and observational advances that may allow the community to better discern if and to what extent a record of disc dispersal by photoevaporation is preserved in observed giant planet separations (and by extension, whether photoevaporation is the primary agent driving disc dispersal). 

\section{Model}
\label{sec:mod}
\citet[hereafter ER15]{Ercolano2015} use a 1D viscous evolution code, $\textsc{SPOCK}$, to model a giant planet undergoing Type II migration in a disc subject to a pure EUV or X-ray dominated photoevaporative wind. The modeling approach emulates that in \citet[hereafter AP12]{Alexander2012a}, which is based on that of \citet{Alexander2009}. We replicate ER15 simulations using $\textsc{SPOCK}$ and extend the code to include an FUV dominated wind. See those works for a full model description; we summarize the key features.

The planet-disc system evolves according to 
\begin{equation}
\frac{\partial \Sigma}{\partial t} = \frac{1}{r} \frac{\partial}{\partial r} \bigg[ 3r^{1/2} \frac{\partial}{\partial r}\big(\nu \Sigma r^{1/2}\big) - \frac{2 \Lambda \Sigma r^{3/2}}{(G M_*)^{1/2}}\bigg] - \dot{\Sigma}_{\rm wind}(r,t),
\label{eq:evo}
\end{equation}
where the first term on the right-hand side describes the viscous evolution of the disc
\citep{Lynden-Bell1974}, the second term the migration of the planet due to torques from the disc \citep[e.g.,][]{Lin1986}, and the last term the mass loss due to photoevaporation \citep[e.g.,][]{Clarke2001}. $\Sigma$ is the gas disc surface density, $r$ the radial distance from the star in the disc midplane, $G$ the gravitational constant, $M_*$ the stellar mass, and {\sd} the radial photoevaporation profile. $\nu$ is the kinematic viscosity of the disc, which we prescribe as $\nu=\alpha c_{\rm s} H$ \citep{Shakura1973}, where $H$ is the disc scale height. 

$\Lambda$ is the rate of specific angular momentum transfer from the planet to the disc. We use the modification by \citet{Armitage2002} to the form proposed by \citet{Lin1986},
\begin{equation}
\Lambda(r,a) = \left\{
        \begin{array}{ll}
            \frac{-q^2 G M_*}{2r}\bigg(\frac{r}{\Delta_p}\bigg)^4 & \quad r<a \\
            \frac{q^2 G M_*}{2r}\bigg(\frac{a}{\Delta_p}\bigg)^4 & \quad r>a,
        \end{array}
    \right.
\end{equation}
where $q$ is the mass ratio between the planet and the star, $a$ is the
semimajor axis of the planet's orbit (assumed to be circular), and $\Delta_p = {\rm max}(H, |r-a|)$. 

The planet is simply inserted into the disc after a formation time $t_{\rm p,0}$; we do not model the planet's formation. Due to torques from the disc immediately interior and exterior to its orbit, the planet begins moving inward at the Type II migration rate \citep[e.g.,][]{Kley2012a},
\begin{equation}
\frac{da}{dt} = \bigg(\frac{a}{GM_*}\bigg)^{1/2} \bigg(\frac{4\pi}{M_{\rm p}}\bigg) \int_{r_{\rm in}}^{r_{\rm out}} r\Lambda \Sigma\ dr,
\label{eq:mig}
\end{equation}
opening its own circumplanetary gap. {\map} is the planet's mass, and the planet accretes gas as it migrates. We performed tests holding a planet at its insert location for a few orbits to prevent rapid initial migration before the disc has had time to adjust to its presence, but saw negligible effect on final orbital separations.

We discretize Equation~\ref{eq:evo} on a grid of 1000 cells equispaced in $r^{1/2}$ between $0.04 - 10^4$ AU (increased to 4000 cells once a planet is inserted in the disc to ensure numerical convergence of the planetary torques and thus migration rate). We assume a disc temperature structure $T \propto r^{-1/2}$, with $T = 2100$ K and 4 K at the inner and outer boundaries. Although throughout the text we refer only to the value of the viscosity coefficient $\alpha$, note that the physical quantity in our equations for the gas is the kinematic viscosity $\nu$; the values of $\alpha$ we use are therefore degenerate with our disc temperatures. To integrate the viscous term in Equation~\ref{eq:evo} we perform a change of variables to recast the equation into a diffusion equation \citep{Pringle1986}. The planet torque is treated as an advection term and computed after the diffusive term using operator splitting. We use the \citet{vanLeer77} method to reconstruct the surface density at the cell boundaries. The photoevaporation term is integrated by removing a fixed amount of mass from each cell at every timestep. To prevent numerical problems we use a floor surface density of $10^{-8}$ g cm$^{-2}$. We limit the maximum torque close to the planet for computational reasons; this has no consequence on the orbital migration rate.

\subsection{Planetary accretion}
To treat mass flow across a gap induced by the planet we use the prescription of \citet{Veras2004},
\begin{equation}
\label{eq:epsilon}
\frac{\epsilon (M_{\rm p})}{\epsilon_{\rm max}} = 1.67 \bigg(\frac{M_{\rm p}}{1\ {\rm M_J}} \bigg)^{1/3} {\rm exp} \bigg(-\frac{M_{\rm p}}{1.5\ {\rm M_J}}\bigg) + 0.04.
\end{equation}
The efficiency $\epsilon(M_{\rm p})$ is the ratio of the accretion rate onto the planet to the accretion rate in a disc without a planet. The accretion rate onto the planet is then related to the disc accretion rate by $\dot{M}_{\rm p}=\epsilon(M_{\rm p})\dot{M}_{\rm disc}$. Consequently we compute the mass accretion rate through the planet's gap  as $\dot{M}_{\rm inner} = \dot{M}_{\rm p}/(1+\epsilon)$. Note that with this prescription $\dot{M}_{\rm inner} + \dot{M}_{\rm p} \neq \dot{M}_{\rm disc}$, though we follow it for consistency in comparing our results with those in AP12 and ER15.
To apply this mass leakage through the gap computationally we set the cells outside the planet's orbit to the floor value until we have removed the mass that is flowing in the timestep. We then use the prescribed rate $\dot{M}_{\rm p}$ to increase the planet mass and \lq{}unload\rq{} the mass coming from the rate $\dot{M}_{\rm inner}$ onto the first cell inside the planet's orbit.

\subsection{Photoevaporation}
\label{sec:winds}
To test \textit{only the effect of the photoevaporative mass loss profile} (FUV, EUV or X-ray driven) on gas giant migration, we use a single disc model for all simulations, that of an X-ray irradiated disc in ER15. Its parameters are summarized in Table~\ref{tab:models}; it uses a Shakura-Sunyaev viscosity parameter $\alpha \approx 7.5 \times 10^{-4}$ in a 75 AU disc with scaling radius $r_1 = 18$ AU and initial mass 0.07 {\ms} around a 0.7 {\ms} T Tauri star. The initial disc size is set by exponentially tapering the self-similar solution to the diffusion equation for the disc surface density \citep{Lynden-Bell1974} at the scaling radius, $r_1$, and 75 AU is the location at which $\Sigma = 1$ g cm$^{-2}$ along this exponential decline.

Radial photoevaporative profiles, {\sd}, for each energetic regime have two epochs: before and after photoevaporation clears a hole in the disc. These are shown in Figure~\ref{fig:discs}(a) -- (b). Each profile is static until a hole opens, at which point the functional form is altered (as described below) to concentrate mass loss at the inner edge of the outer disc $r_{\rm in}$ that is now exposed directly to stellar irradiation. In the EUV and X-ray regimes the profile's integrated mass loss rate may also change.    

While the X-ray and FUV profiles are thought to drive integrated mass loss rates a factor of $\sim 100$ greater than an EUV profile, here we draw the integrated mass loss rate for all three regimes from the same distribution to test only the effect of the difference between radial photoevaporative profiles on giant migration. We may be modeling unrealistically high EUV mass loss rates, but choose this for the purpose of direct comparison across profiles. Note that the absolute mass loss rates and disc viscosity parameters are degenerate in their effect on the disc lifetime, allowing similar simulated disc lifetimes over a wide (greater than an order of magnitude) range of mass loss rates. Moreover in the EUV and X-ray cases the mass loss profile is largely independent of the total luminosity incident on the disc (which is a difficult quantity to measure). Consequently we feel a comparison between the profiles only (rather than unique disc models to accompany each profile) more robustly tests the differences in how each heating mechanism (that in the FUV, EUV and X-ray dominated regimes) affects the disc evolution and migrating planets. 

\begin{table*}
\caption{Model star and disc properties. The standard deviation $\sigma$ assumes a Gaussian distribution.}
\begin{tabular}{l c c}
    \hline
    Star mass [\ms] & 0.7\\
    Initial disc mass [\ms]& $0.07$\\ 
    Disc mass loss rate, {\md} [{\ms} yr$^{-1}$] & median $\approx 7.5 \times 10^{-9},\ \sigma \approx 2.6 \times 10^{-9}$\\
    Viscosity coefficient, $\alpha$ & $\approx 7 \times 10^{-4}$\\
    Scaling radius, $r_1$ [AU] & 18\\
    Viscous time at $r_1$, $t_\nu$ [yr] & $7 \times 10^5$\\
    Disc aspect ratio at $r_1$, $H/r$ & 0.1\\
    \hline
\end{tabular}
\label{tab:models} 
\end{table*}

\subsubsection{X-ray profile}
\label{sec:x}
We use the X-ray dominated photoevaporative mass loss profile shown in Figure~\ref{fig:discs}(a) used in \citet{Owen2010}, \citet{Owen2011} and \citet[see their Equation B2]{Owen2012} that includes a secondary EUV component and is derived using a hydrodynamic solution for the wind. We draw integrated photoevaporative mass loss rates using the distribution of X-ray luminosities in the Taurus cluster \citep[see \citealp{Owen2011} Figure 1]{Gudel2007} that approximately reproduces the observed scatter in stellar accretion rates and thus disc lifetimes \citep{Owen2011}, with the relation between luminosity and disc mass loss rate as in \citet{Owen2012} for a disc without a hole,
\begin{equation}
\label{eq:lx}
\dot{M}_{\rm wind} = 6.25 \times 10^{-9}\ \Big(\frac{M_*}{1\ {\rm M_\odot}}\Big)^{-0.068}\ \Big(\frac{L_{\rm X}}{10^{30}\ {\rm erg\ s^{-1}}}\Big)^{1.14}\ {\rm M_\odot\ yr^{-1}}.
\end{equation}
The median $L_{\rm X} = 1.2 \times 10^{30}$ erg {\s} yields an integrated mass loss rate $\dot{M}_{\rm wind} = 7.5 \times 10^{-9}$ {\ms} {\yr} with standard deviation $\sigma = 2.6 \times 10^{-9}$ {\ms} {\yr} (assuming a Gaussian distribution).

Assuming an X-ray penetration depth of $10^{22}$ cm$^{-2}$ \citep{Ercolano2009}, in the absence of a planet we switch to a second epoch in the photoevaporative profile once the hydrogen column density in the disc midplane is below this level out to 1.7 AU \citep[see Equation B3 of][]{Owen2012}. Photoevaporation has now cleared a hole in the disc, and the mass loss profile dynamically evolves to concentrate mass loss at the inner edge of the outer disc, $r_{\rm in}$ \citep[shown in Figure~\ref{fig:discs}(b); see Equation B5 of][]{Owen2012}. The integrated mass loss rate may also change as the mass loss profile's peak tracks $r_{\rm in}$.

\subsubsection{EUV profile}
The EUV {\sd} profile is derived from hydrodynamic simulations, and we switch from its first (\lq{}diffuse\rq{}) epoch \citep[][see Equation A1]{Alexander2007} to a second (\lq{}direct\rq{}), dynamic form \citep[Equation A5 of][]{Alexander2007} that concentrates mass loss at $r_{\rm in}$ once the EUVs penetrate a column density of $10^{18}$ cm$^{-2}$ \citep{Ercolano2009}. The disc lifetime is insensitive to variation of the threshold column density over $\pm\ {\rm 2}$ orders of magnitude. As in the X-ray case, the integrated mass loss rate may change as the profile evolves with $r_{\rm in}$.

\subsubsection{FUV profile}
In the first epoch we use a static FUV mass loss profile that is an average of the time-dependent FUV dominated model (with secondary X-ray and EUV contributions) in \citet{Gorti2009}. These authors used a 1+1D hydrostatic equilibrium model to determine the vertical gas temperature and density profiles in the disc, calculating $\dot{\Sigma}_{\rm wind}(r,z) \propto \rho_{\rm b} c_{\rm s}$ and taking $\dot{\Sigma}_{\rm wind}(r)  = {\rm max}(\dot{\Sigma}_{\rm wind}(r,z))$ at each radius in the disc, where $\rho_{\rm b}$ is the gas density at the base of the flow.
The FUV luminosity is thought to be a consequence primarily of stellar accretion and would thus decline over time; in \citet{Gorti2009} this decrease in {\md} is stable to within a factor of a few over the disc lifetime. 

We switch the profile to its second epoch when the FUV photons penetrate a column density of $10^{22}$ cm$^{-2}$ (as in the X-ray case). The profile interior to the trough at 18 AU (Figure~\ref{fig:discs}(a)) is then continuously shifted so that its maximum (initially at 7.3 AU) coincides with $r_{\rm in}$ while keeping the integrated {\md} equal to the fixed value used in the first epoch. We do this for lack of a $\dot{\Sigma}$ prescription derived using a hydrodynamic solution for the FUV wind. The profile exterior to 18 AU is left unaltered because physically the photoevaporative hole ought only to affect the disc structure in its vicinity.

\subsection{Population synthesis}
\label{sec:popsyn}
To construct a population synthesis for disc-planet systems in each of the three photoevaporation regimes we randomly sample the initial photoevaporative mass loss rate from the distribution in Section~\ref{sec:x}, initial planet mass from a uniform distribution $0.5 \leq M_{\rm p} \leq 5.0\ {\rm M_J}$ and planet formation time from a uniform distribution $0.25 \ {\rm Myr} \leq t_{\rm p,0} \leq t_{\rm c}$, where $t_{\rm c}$ is the time of disc clearing \citep{Clarke2001, Ruden2004},
\begin{equation}
t_{\rm c} = \frac{t_\nu}{3}\bigg(\frac{3M_{\rm d}}{2t_\nu \dot{M}_{\rm wind}}\bigg)^{2/3}.
 \label{eq:tc}
\end{equation}
$t_\nu = (\alpha \Omega)^{-1} (H/r)^{-2}$ is the viscous time, $\Omega$ the Keplerian angular velocity, and $M_{\rm d}$ the disc mass. The planet is always inserted at 5 AU, and 0.25 Myr is chosen as an arbitrary lower bound on formation time under a simple assumption for some minimum time required to form a gas giant. 

We perform 1000 realizations of the model using each the FUV, EUV and X-ray photoevaporative mass loss profiles. A simulation ends when the planet's orbital separation $a \leq 0.15$ AU (below which we do not track it as we do not attempt to model the planet's interaction with the disc's magnetospheric cavity) or the disc is sufficiently dispersed, which we define as the time at which the inner rim of the outer disc (following the opening of a hole) $r_{\rm in} \geq {\rm min}(1.5 a, 10\ {\rm AU})$. The planet has been stalled in its migration for several orbits before either dispersal condition is met.

\subsection{Model summary}
Procedural steps of the model (in greater detail above) are:\\
1) Begin the 1D simulation of a viscously evolving disc by Equation~\ref{eq:evo}, with disc and star parameters as given in Table~\ref{tab:models}. Irradiate the disc with either a pure EUV, X-ray dominated or FUV dominated stellar flux, drawing the total photoevaporative mass loss rate from a distribution of stellar X-ray luminosities and the conversion in Equation~\ref{eq:lx}, centered at $\approx 7.5 \times 10^{-9}$ {\ms} {\yr} and spanning $\sim 10^{-10} - 10^{-7}$ {\ms} {\yr}. \\
2) Insert a formed planet at a random time between 0.25 Myr and the time of disc clearing (Equation~\ref{eq:tc}) whose mass is sampled from a uniform distribution over the range $0.5\ {\rm M_J} \leq M_{\rm p} \leq 5.0\ {\rm M_J}$. At this time, increase the grid from 1000 to 4000 cells.\\ 
3) The planet moves inward by Type II migration (Equation~\ref{eq:mig}) and accretes mass as gas flows across its gap by Equation~\ref{eq:epsilon}, while the disc viscously evolves and photoevaporates.\\
4) After $\approx$ few Myr, photoevaporation dominates viscous evolution and opens a hole by penetrating a hydrogen column density radially outward in the disc midplane of $10^{18}(10^{22})\ {\rm cm^{-2}}$ in the EUV (FUV or X-ray) regime. This exposes the outer disc directly to stellar irradiation; from this time onward, the radial photoevaporative profile concentrates mass loss at the hole radius $r_{\rm in}$. In the EUV and X-ray regimes the profile's integrated mass loss rate may also change.\\
5) End the simulation when either the disc surface density is $\leq 10^{-8}$ g cm$^{-2}$ at all radial grid points out to 10 AU or out to 1.5x the planet's separation, or the migrating planet reaches a separation $\leq 0.15$ AU (below which we do not attempt to model its interaction with the magnetospheric cavity).\\ 
6) Conduct 1000 simulations for each of the three photoevaporative profiles, varying only the photoevaporative mass loss rate, planet formation time and initial planet mass.

\section{Results \& discussion}
\label{sec:res}

\subsection{Illustrative cases}
\label{sec:sample_cases}
Figure~\ref{fig:discs}(d) -- (f) show the evolution of the disc surface density $\Sigma$ in the absence of a planet using the X-ray, EUV and FUV radial mass loss profiles respectively for the median mass loss rate {\md} $= 7.5 \times 10^{-9}\ {\rm M_\odot\ yr^{-1}}$. The differing {\sd} profiles result in the unique time and location at which a photoevaporative gap is opened across models, as well as the timescale over which the gap becomes a hole and the disc is dispersed. A gap is first opened near the EUV profile's peak (0.8 AU) at 2.0 Myr, near the X-ray profile's peak of 1.7 AU at 2.4 Myr, and near 4 AU with the FUV profile (whose broad peak extends from 3 - 10 AU) at 3.0 Myr. The disc is subsequently dispersed out to 10 AU in 140 kyr with the EUV, a factor of 2.5 longer with the X-ray and of 4.5 with the FUV profile. The photoevaporative profile strongly influences the separation at which planets in our population syntheses are stalled in their migration largely as a consequence of this hierarchy of dispersal times, $t_{\rm disc,\ EUV} < t_{\rm disc,\ X-ray} < t_{\rm disc,\ FUV}$. Section~\ref{sec:migra_photoevap} discusses trends in the distinct effect each photoevaporative profile has on migrating planets of different masses and the physical processes underlying these trends.

\subsubsection{Interplay of migration, mass transfer through the circumplanetary gap and photoevaporation}
\label{sec:migra_photoevap}
\begin{figure*}
	\includegraphics[width=\textwidth]{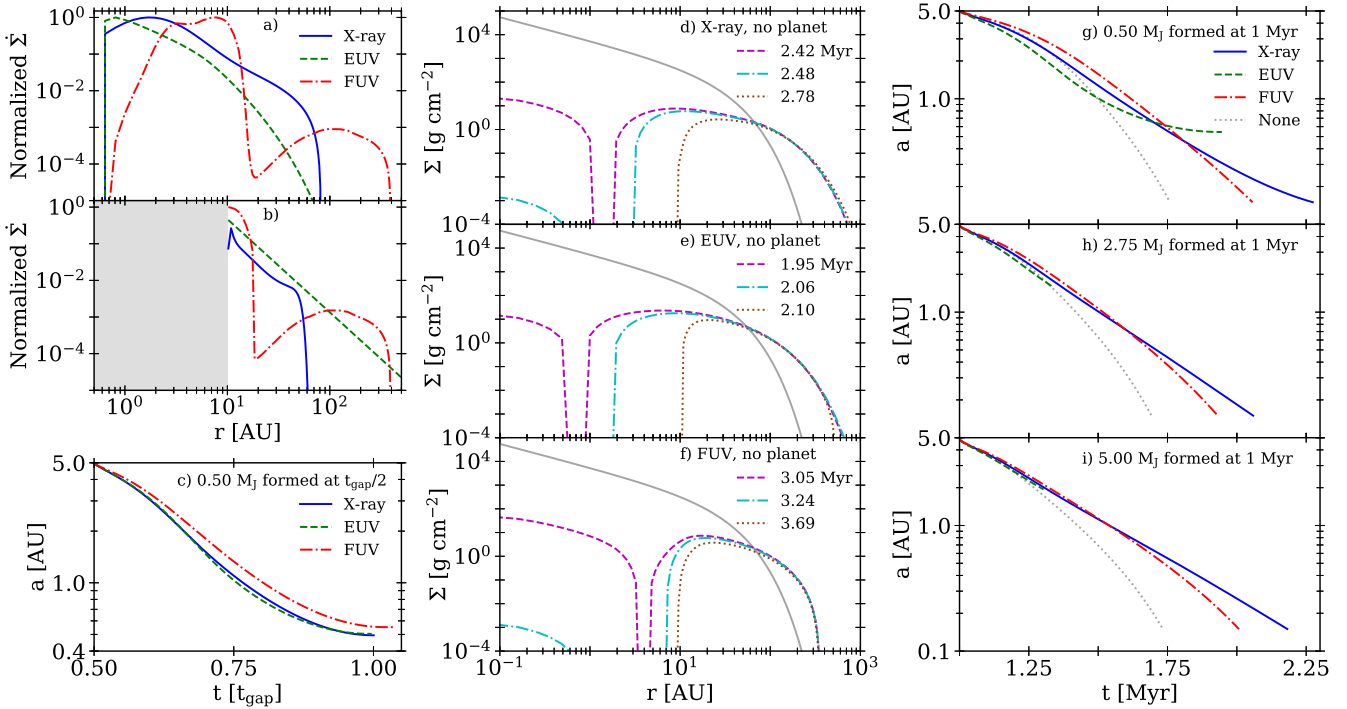}
	    \caption{a) Normalized radial photoevaporative mass loss profiles for X-ray, pure EUV and FUV dominated winds. These profiles are static until photoevaporation clears a hole in the disc. 
The X-ray and FUV dominated profiles' integrated mass loss rates are predicted to be 2 orders of magnitude larger than that of the EUV, though in this study we draw mass loss rates for all three regimes from the same distribution to compare only the effect of the radial profile.
b) Normalized radial photoevaporative mass loss profiles after photoevaporation clears a hole that, shown here, extends out to 10 AU in the disc (shaded region). All profiles are dynamic once a hole opens, including variation of the integrated mass loss rate in the X-ray dominated and pure EUV cases.
c) Separation over time of a 0.50 {\mj} planet inserted in the discs of panels (d) -- (f) at 5 AU at 50\% of the time taken to open a photoevaporative gap in those discs: 1.210 Myr in the X-ray case, 0.975 Myr in the EUV and 1.525 Myr in the FUV.  d) Evolution of the disc surface density over time assuming a mass loss rate {\md} $= 10^{-8}\ {\rm M_\odot\ yr^{-1}}$ integrated over an X-ray dominated photoevaporation profile: the disc at time $t_0$ (gray), when photoevaporation first opens a gap in the disc ($t = 2.42$ Myr), at the time this gap becomes a hole ($t = 2.48$ Myr), and at disc dispersal ($t = 2.78$ Myr). No planet is included in the simulation. The decline in surface density across the disc from $t_0$ to the time of gap opening is predominantly due to the disc expanding as it evolves viscously and conserves angular momentum. e) -- f) As in (d) for the EUV and FUV dominated profiles, using the same integrated mass loss rate. g) Separation over time of a 0.50 {\mj} planet inserted in the discs of panels (d) -- (f) at 5 AU at 1 Myr. Also shown is a planet inserted in an equivalent disc without photoevaporation. h) -- i) As in (g) for a 2.75 and 5.00 {\mj} planet. Trends across photoevaporation regimes and planet mass are discussed in Sections~\ref{sec:sample_cases} and~\ref{sec:migra_photoevap}.}
    \label{fig:discs}
\end{figure*}

One instance of the interaction between photoevaporation and our prescription for giant migration is demonstrated in the example cases of Figure~\ref{fig:discs}(g) -- (i). These panels place either a 0.50, 2.75 or 5.00 {\mj} planet (the minimum, mean and maximum initial planet masses used in our simulations) at 5 AU at 1 Myr in a disc subject to either an X-ray, EUV or FUV driven photoevaporative wind with a mass loss rate {\md} $= 10^{-8}\ {\rm M_\odot\ yr^{-1}}$. Additionally for reference the panels show the planet inserted in an equivalent disc without photoevaporation.

One may expect in the Type II migration regime that a less massive giant should migrate faster, yet in the X-ray and FUV cases of Figure~\ref{fig:discs}(g) -- (i) the 2.75 {\mj} body reaches our 0.15 AU boundary before its 0.50 and 5.00 {\mj} counterparts. This is a result of the rate of mass transfer across the circumplanetary gap (from the outer to inner disc) and the way in which that affects photoevaporation to complicate the Type II scaling. A heavier planet has a lower accretion efficiency under the parameterization we impose (sensitivity to which is discussed in Section~\ref{sec:unc}), more strongly impeding mass flow across its gap relative to a lighter planet. This produces an excess of mass exterior to the heavier planet's orbit to drive faster inward migration. Following planet formation (or in our model, insertion of a formed planet in the disc), the body initially causes a dearth of mass exterior to its gap, slowing its migration. Mass transfer across the circumplanetary gap is then reduced, impairing replenishment of the inner disc as it drains onto the star. Because the body is migrating sub-viscously (all planets in Figure~\ref{fig:discs}(g) --(i) have migration rates less than the local viscous rate, $da/dt < v_{\rm r} = -3/2\ \nu/r$ as a consequence of the object's mass exceeding the local disc mass), the outer disc eventually refills this underdensity exterior to the planet's gap, pushing the body inward at a resumed faster rate. This takes longer to occur for the lighter planet because of its high accretion efficiency (high rate of mass transfer across the circumplanetary gap), delaying a buildup of mass exterior to its orbit and more quickly resupplying mass to the inner disc. 

Consequently photoevaporation is able to drive mass loss from the lowered surface density disc $exterior$ to a less massive planet, causing an extended period of slowed migration \citep{Rosotti2013}. By contrast more massive giants' greater suppression of mass transfer from the outer to inner disc yields an earlier buildup of mass exterior to their orbits, more quickly ending a period of slowed migration. The heavier planets' low accretion efficiency also causes a reduced disc surface density $interior$ to their orbits, resulting in a weaker outward torque exerted on those bodies as it allows photoevaporation to more easily drive mass loss interior to the planet. In cases of sufficiently massive planets formed sufficiently late in the disc lifetime such as the 2.75 and 5.00 {\mj} bodies in the EUV cases of Figure~\ref{fig:discs}(h) -- (i), this reduced disc surface density interior to the planet can spur a photoevaporative gap opening in the inner disc by effectively allowing photoevaporation to extend the circumplanetary gap inward. Rapid subsequent disc dispersal follows, stalling the planet at larger separations than in the FUV and X-ray regimes. Note the planet in these cases is inducing disc dispersal $\approx 800$ kyr earlier than would occur solely due to photoevaporation in the absence of a planet (as in Figure~\ref{fig:discs}(e)). This process is common in our simulations and is discussed further in Section~\ref{sec:trends}. 

The initially slower migration in the FUV case, evident especially in Figure~\ref{fig:discs}(g), is due to the FUV photoevaporative profile's comparatively high(low) disc mass loss at 5(1) AU, reducing the inward torque exterior to the planet while leaving the outward torque relatively unhindered. In spite of this the bodies in an FUV irradiated disc migrate to our 0.15 AU cutoff faster than in the EUV or X-ray regimes primarily as a result of 1 Myr being comparatively earlier in the FUV disc lifetime (recall $t_{\rm disc,\ EUV} < t_{\rm disc,\ X-ray} < t_{\rm disc,\ FUV}$).   

For a direct comparison of the photoevaporative profiles' differing effects on a migrating planet at an equivalent time in the disc's evolution (i.e., when the surface density profile is comparable for discs subject to X-ray, EUV or FUV irradiation), Figure~\ref{fig:discs}(c) inserts a 0.50 {\mj} planet in the discs of panels (d) -- (f) at 50\% of those simulations' photoevaporative gap opening times, e.g., 1.21 Myr in the X-ray case. The discrepancy between planet migration rates in each photoevaporative regime and thus the separations at which the bodies are stalled is due to the photoevaporative profiles' differing shapes, primarily the location and width of the peak. In our population syntheses we draw from the same distribution of stellar luminosities in each photoevaporative regime (giving results for which Figure~\ref{fig:discs}(g) -- (i) are indicative) rather than the same spread in disc lifetimes (for which Figure~\ref{fig:discs}(c) would be indicative) because we treat the time taken to disperse the disc as an intrinsic property of the photoevaporative profile.

Note the trends described here are sensitive to the time at which the planet is formed and the strength of the photoevaporative wind. For example, in general across our population synthesis results less massive planets do migrate faster and thus further as discussed in Sections~\ref{sec:charac} and~\ref{sec:trends}. The example cases here are meant to elucidate some of the more subtle physical processes in the interaction between migration and disc photoevaporation in our model that produce characteristic features in the separation distribution of gas giants in discs subject to each photoevaporative regime.

\subsection{Characteristic features of gas giant populations subject to FUV, EUV or X-ray dominated disc photoevaporation}
\label{sec:charac}
\begin{figure*}
	\includegraphics[width=\textwidth]{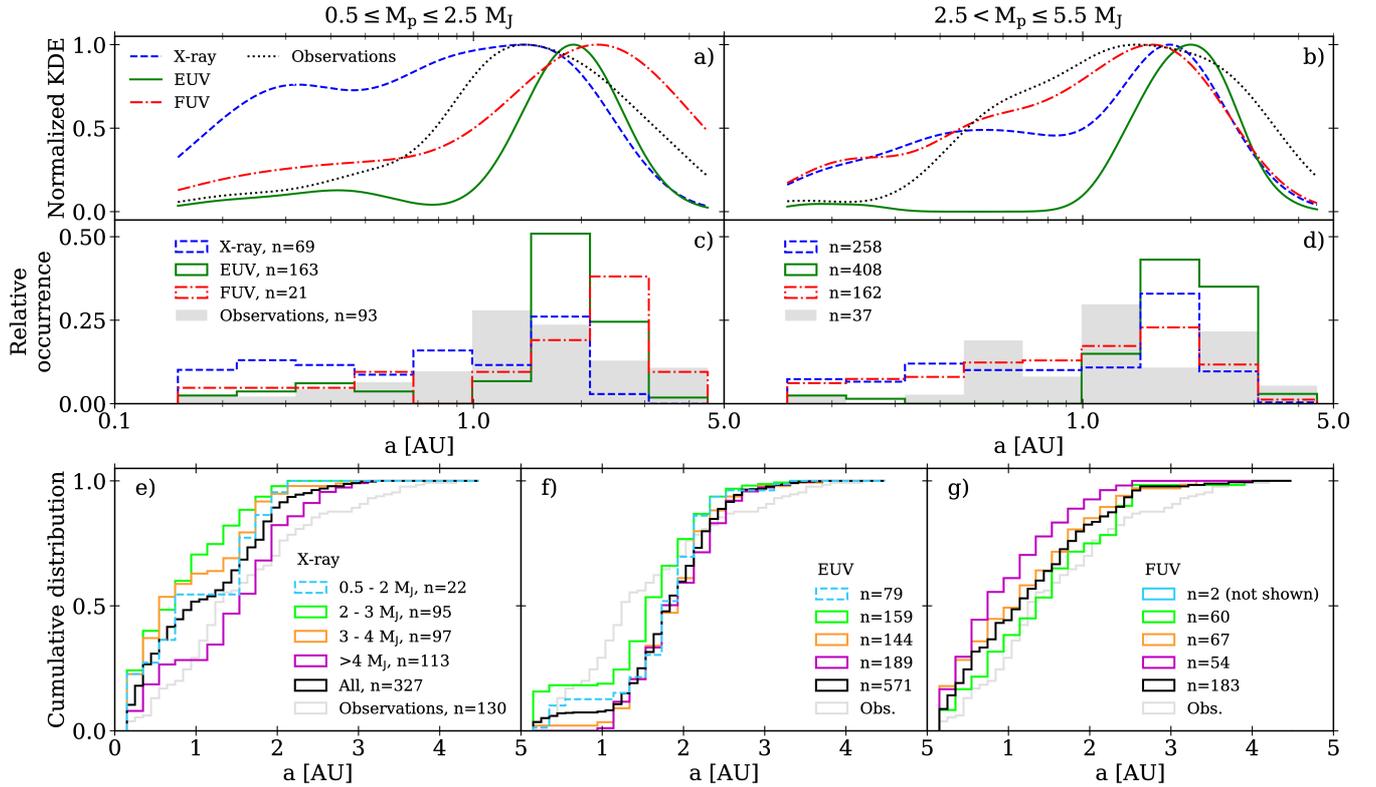}
    \caption{a) Gaussian kernel density estimate for the separation at which planets of $final$ mass $0.5\ {\rm M_J} \leq M_{\rm p} \leq 2.5\ {\rm M_J}$ are halted in their migration under either an FUV, EUV or X-ray driven photoevaporative mass loss profile. Radial velocity detections in the exoplanets.org catalog \citep{Han2014} at the time of publication for single planet systems with minimum masses in the range $0.5\ {\rm M_J} \leq M_{\rm p}\sin{i} \leq 5.5\ {\rm M_J}$ (the latter being the largest final planet mass in our simulations), separations between $0.15 - 4.50$ AU and host star masses $0.75 - 1.50$ {\ms} are shown for comparison (note the sample is not complete out to 5 AU).
We omit simulated planets whose final separation $a \leq 0.15$ AU as described in Section~\ref{sec:charac} and end the distributions at 4.5 AU for comparison with AP12 and ER15. Note the former condition removes  $\approx 41\%$, 66\% and 81\% of all simulations under the EUV, X-ray and FUV profiles respectively. The latter condition removes no simulations. b) As in (a) for gas giants with final masses $2.5 < M_{\rm p} \leq 5.5\ {\rm M_J}$. c -- d) As in (a) -- (b), recast in histograms for comparison with AP12 Figure 3 and ER15 Figure 3. e) Cumulative distribution of separations for planets halted in their migration between $0.15 - 4.50$ AU under an X-ray driven photoevaporation profile, binned by final planet mass. The cumulative histogram for all planets (black) uses 0.1 AU bins; all others use 0.2 AU bins. The observational sample's distribution (all masses) is shown for comparison in gray, 0.1 AU bins. f -- g) As in (e) for the EUV and FUV profiles. (e) -- (g) can be compared with AP12 Figure 1. Characteristic features in the distributions under each photoevaporative regime, as well as similarities and disparities between the observational sample and model results, are discussed in Section~\ref{sec:charac}.}
    \label{fig:hist}
\end{figure*}

\textit{The location and width of the peak in the photoevaporative mass loss profiles we use are the principal variables determining our results, largely as a consequence of the different time each takes to disperse an equivalent disc} (see Figure~\ref{fig:discs}(d) -- (f)). A consequence of these unique profiles and their effects on migrating planets, the separation distributions in Figure~\ref{fig:hist} show where giants are halted in their migration under each photoevaporative regime. This figure casts results in terms of final planet mass and omits bodies with final separations $a \leq 0.15$ AU, as we do not attempt to model planet interactions with the stellar magnetospheric cavity in the innermost disc. \textit{This condition removes 41\%, 66\% and 81\% of planets in the EUV, X-ray and FUV models respectively}, and the vast majority of planets reaching this 0.15 AU cutoff are still migrating quickly, suggesting they would either be lost onto the host star or halted at periods $\lesssim 0.1$ AU.

While this result is sensitive to many poorly constrained parameters, if as a hypothetical a significant fraction of young gas giants do fall onto their star (as hot Jupiters are rare), implications could include a typical disc metals inventory well in excess of the minimum mass solar nebula and a non-trivial fraction of polluted T Tauri stars. However regarding the latter, \citet{Laughlin1997} explore stellar metallicity enhancement due to accretion of gas giants with Jupiter metallicity ($Z_{\rm p} \approx 0.1$) and find that for pre-solar T Tauri stars, the effect is severely diluted because the planetary material is mixed throughout the fully convective host. They do find that the effect becomes significant (stellar metallicity enhancement $\Delta Z_* \gtrsim 0.005$) for $\approx {1.5 - 2\ \rm M_\odot}$ young stellar objects as a consequence of their thin convective zones.

Figure~\ref{fig:hist}(a) -- (b) show Gaussian kernel density estimates for the separation distribution in each photoevaporation regime. We split the data into two mass bins to separate the more massive planets from the lower mass set that comprises the bulk of the observational sample, as well as for comparison with AP12 and ER15. The same data are binned in histograms in Figure~\ref{fig:hist}(c) -- (d) for a more direct comparison with those works. Note our distributions are not fully converged (with respect to sample size) on their shapes shown here, but the location of the global maximum in each is robust, and the characteristics we discuss are insensitive to these variations. We judged this by comparing the distributions obtained with randomly drawn sets of simulations, which showed consistency in the features listed below for sets of $\gtrsim 500$ runs (the population synthesis for each photoevaporative profile contained 1000 total runs). We find the most prominent features of our distributions under an FUV, EUV or X-ray profile are:\\

\noindent - FUV: Few surviving low mass ($\lesssim 2$ {\mj}) giants. Approximately 81\% of planets with initial mass $0.5\ {\rm M_J} \leq M_{\rm p,0} \leq 2.0\ {\rm M_J}$ migrate interior to our 0.15 AU cutoff, compared with 66\% in the X-ray and 46\% in the EUV simulations. Of the planets with initial mass $\leq 2$ {\mj} that are halted in their migration, 97\% in the FUV have a final mass $> 2$ {\mj} (79\% in the X-ray and 56\% in the EUV). The paucity of low mass gas giants produced in an FUV irradiated disc is primarily a consequence of: a much broader mass loss profile peak than in the EUV or X-ray, requiring a comparatively long time to open a photoevaporative gap in the disc ((Figure~\ref{fig:discs}(f); note this also causes a smaller fraction of $> 2$ {\mj} bodies to survive in the FUV than in the EUV and X-ray regimes); and the profile's peak location at larger radii in the disc than the EUV or X-ray, resulting in a relative inefficiency at driving mass loss interior to 2 AU (Figure~\ref{fig:discs}(a)). These features (typically) allow gas giants that migrate interior to the FUV peak's inner edge to progress in a relatively unperturbed disc, giving them more time to migrate (and grow as they do so) before photoevaporation can become effective (with the planet's aid) to disperse the disc and stall the body. 
\\

\noindent - EUV: A high concentration of $\lesssim 3$ {\mj} planets between $\approx 1.5 - 2$ AU. The peak in our model distribution of giant separations is just exterior to the peak in the EUV mass loss profile, a consequence of the circumplanetary gap spurring photoevaporative dispersal as the planet approaches the mass loss profile's peak. This trend is discussed further in Section~\ref{sec:trends} and is also responsible for the high percentage of planets in the EUV being stalled in their migration exterior to our 0.15 AU cutoff (59\%, contrasted with 34\% in the X-ray and 19\% in the FUV). While many model and disc parameters carry high uncertainties, we expect a concentration in stalled planets near the photoevaporative profile's peak to be robust in an EUV dominated regime because the profile's sharp peak (Figure~\ref{fig:discs}(a)) permits a planet to migrate in the disc largely unaffected by photoevaporation until the body nears it. We see this hold strongest for $\lesssim 3$ {\mj} giants because beyond this mass the planet's wider circumplanetary gap can spur the opening of a photoevaporative gap sooner (when the planet is at a larger separation).
\\

\noindent - X-ray: A comparative abundance of $\approx 2 - 3.5$ {\mj} planets interior to 0.5 AU. Under an X-ray profile intermediate mass giants can show the fastest migration (as a function of planet mass) in our models as discussed in Section~\ref{sec:migra_photoevap}. Consequently an overabundance of planets in this mass range relative to the EUV and FUV regimes either breach our 0.15 AU boundary or are halted between 0.15 -- 0.50 AU.
\\

Figure~\ref{fig:hist} also includes a simple observational distribution for comparison with our model results, \textit{though because of large uncertainties in disc and planet formation/migration models} (discussed in Section~\ref{sec:unc}), \textit{this comparison is primarily qualitative} and is secondary to a comparison between the FUV, EUV and X-ray model results. We use the set of radial velocity detections in the exoplanets.org catalog \citep{Han2014} at the time of publication to select single planet systems with minimum masses $0.5\ {\rm M_J} \leq M_{\rm p}\sin{i} \leq 5.5\ {\rm M_J}$ (the upper bound being the largest mass to which our simulated planets grow), separations $0.15 \leq a \leq 4.50$ AU and host star masses $0.75 \leq M_* \leq 1.50$ {\ms} to roughly emulate our simulation conditions. Note we do not attempt to correct this simple sample for biases as the intent is not a robust comparison to our model results, but a qualitative sense of whether these results recover the observational distribution's strongest feature, a peak between 1 -- 2 AU that holds across planet mass. The observational sample is largely complete out to $\approx 3$ AU \citetext{\citealp{Cumming2008}; see also \citealp{Winn2018} Figure 3 and \citealp{Dawson2018} Figure 4}, though because of the higher incompleteness between $3 - 5$ AU, this peak may instead be the inner edge of a plateau. Several scenarios for the peak's origin have been investigated \citep{Dawson2013,Petrovich2015a,Antonini2016,Schlaufman2016,Dawson2018}; here we seek only to assess whether our model results hint at disc photoevaporation contributing to the peak's origin, not a robust comparison against other hypotheses.

Our simulated distributions' characteristic features as described above do not match traits in the simple observational sample. At best we can say a rough match of the observational peak with the X-ray maximum in the low mass bin of Figure~\ref{fig:hist}(a) and with the FUV peak in panel (b) may be an indication that disc photoevaporation is playing an appreciable role in shaping planetary system architectures. A more definitive conclusion is precluded by several large uncertainties inherent to viscous disc models, disc photoevaporation and planet migration, as well as the extent to which dynamic interactions after disc dispersal affect orbital configurations. Note for example that AP12 find a closer alignment of the EUV distribution's peak with a similar set of observations using the same photoevaporative profile but different disc model. Collectively these differences between observations and our simulated distributions, as well as the question of our results' robustness, motivate a discussion of the physical uncertainties to which our findings are most sensitive in Section~\ref{sec:unc}.

\subsection{Trends underlying model distributions}
\label{sec:trends}
\begin{figure*}
	\includegraphics[width=\textwidth]{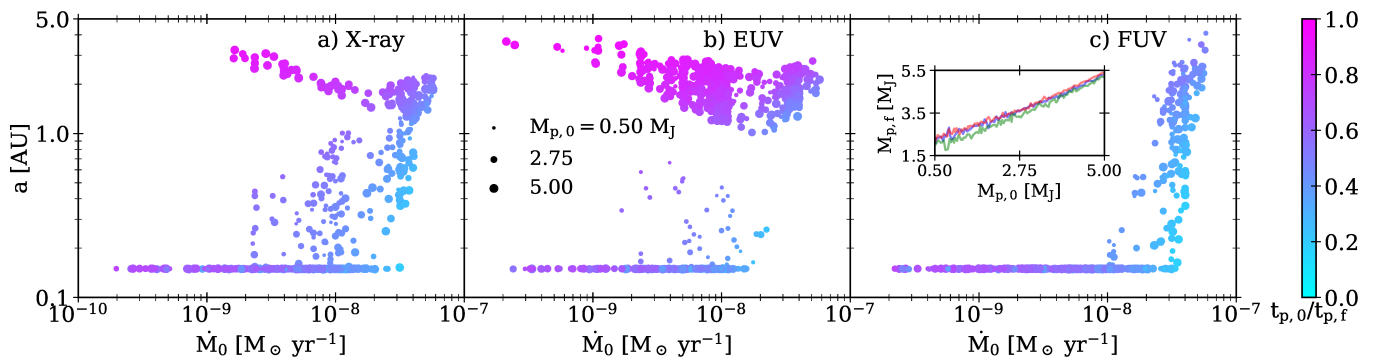}
    \caption{a) For 1000 realizations of the model using an X-ray dominated photoevaporation profile, planet separation at the time of disc dispersal as a function of the initial photoevaporative mass loss rate integrated across the disc. This mass loss rate is constant until photoevaporation clears a hole in the disc, i.e., for most of the disc lifetime (Section~\ref{sec:winds}). Points are colored by the planet's formation time $t_{\rm p,0}$ normalized to the time at which it is stopped in its migration $t_{\rm p,f}$ (the disc dispersal time), as shown in the colorbar. Point size is scaled by $initial$ planet mass (planets accrete mass as they migrate), with example scalings shown in (b). The inset of (c) shows average final planet mass $\rm{M_{p,f}}$ as a function of initial mass for each profile (X-ray, blue; EUV; green; FUV, red), with the spread about these averages decreasing at higher $\rm{M_{p,0}}$.
b -- c) As in (a) for the pure EUV and FUV dominated profiles. All planets are inserted (\lq{}formed\rq{}) at 5 AU; those migrating interior to 0.15 AU are no longer tracked (Section~\ref{sec:popsyn}), producing the artifact at that separation. Note this feature contains 66\%, 41\% and 81\% of of all simulations in panels (a), (b) and (c) respectively. 
Overarching trends and those specific to each profile are discussed in Section~\ref{sec:charac}.}
    \label{fig:form_time}
\end{figure*}

To give greater insight into how our model features emerge, here we motivate the trends in our results, both those common across photoevaporative regimes and those unique to each regime. Figure~\ref{fig:form_time} recasts results in terms of initial planet mass (as opposed to final planet mass in Figure~\ref{fig:hist}), showing the separation at which the body is halted in its migration as a function of the initial photoevaporative disc mass loss rate (once a hole is cleared in the disc this rate can vary; Section~\ref{sec:winds}). Points are scaled in size by initial planet mass and colored by planet formation time $t_{\rm p,0}$ scaled to the time $t_{\rm p,f}$ at which disc dispersal stops the planet's migration (once the planet's position changes by $< 0.01\%$ for 100 orbits). 

In Figure~\ref{fig:form_time} the stratification in $t_{\rm p,0} / t_{\rm p,f}$ at a given $\dot{M}_0$ tracks how depleted the disc is at the time of planet formation; on average, planets inserted later in the disc lifetime are able to more quickly foster photoevaporative disc dispersal by impairing the rate of mass transfer to the inner disc, consequently stalling the planet at larger separations. The slanted envelope below which all stalled planets lie in the FUV case (seen to a successively weaker extent in the X-ray and EUV) is a result of higher mass loss rates depleting the disc faster; the higher the mass loss rate, the more easily (quickly) an inserted planet can facilitate disc dispersal. This envelope only extends down to $\dot{M}_0 \approx 10^{-8}$ {\ms} yr$^{-1}$ in the FUV because of the photoevaporative profile's broad peak and weak mass loss interior to $\approx 2$ AU (Figure~\ref{fig:discs}(a)). If the disc is not sufficiently depleted near 5 AU at the planet's formation time, the broad FUV profile cannot disperse the disc even with the planet's aid. Once the planet has then migrated far enough for its circumplanetary gap to be interior to $\approx 2$ AU, the body faces minimal resistance to continued inward migration and breaches our 0.15 AU boundary. 

The overall trend (at all separations) toward higher $t_{\rm p,0} / t_{\rm p,f}$ at lower $\dot{M}_0$ is an artifact of our range of planet formation times, $0.25\ {\rm Myr} \leq t_{\rm p,0} \leq t_{\rm c}$, where the time of disc clearing $t_{\rm c} \propto \dot{M}_0^{-2/3}$ (Equation~\ref{eq:tc}). The disc lifetime (for which $t_{\rm c}$ is a proxy) is significantly shorter at high mass loss rates, such that the planet's migration timescale is a larger portion of it. 

The disc lifetime also obeys the trend noted in Section~\ref{sec:sample_cases}, $t_{\rm disc,\ EUV} < t_{\rm disc,\ X-ray} < t_{\rm disc,\ FUV}$. This results in the highest average $t_{\rm p,0} / t_{\rm p,f}$ in the EUV and lowest in the FUV. The comparatively short average disc lifetime in the EUV also underlies the structure in Figure~\ref{fig:form_time}(b) at separations $>1$ AU. Points in this grouping are mostly heavy giants with wide circumplanetary gaps inserted after the disc surface density has decreased substantially due to viscous evolution. As soon as the circumplanetary gap's edge approaches the EUV profile's sharp peak, photoevaporation is able to expand the circumplanetary gap inward and drain the inner disc, with the planet then stalled as the outer disc is quickly eroded by direct stellar irradiation. The slope of this grouping toward higher separations at lower $\dot{M}_0$ represents a transition from a scenario of photoevaporation expanding the circumplanetary gap inward once that gap approaches the mass loss profile's peak to a scenario in which the reduced resupply of the inner disc through the circumplanetary gap is sufficient to allow photoevaporation to clear a photoevaporative gap in the inner disc independent of the circumplanetary gap, stalling the planet at the largest separations. The former interaction scenario also occurs in the X-ray case of Figure~\ref{fig:form_time}(a), though because the X-ray profile is less narrowly peaked than the EUV, the effect is more severe in an EUV regime.

Together these processes yield three general interaction regimes between the migrating planet and photoevaporation that determine the density of points in Figure~\ref{fig:form_time} and result in the characteristics features discussed in Section~\ref{sec:charac}: \\

\noindent - Most often under an FUV or X-ray profile (and for 40\% of the simulations under an EUV profile), the planet migrates quickly enough to reach our 0.15 AU cutoff before photoevaporation becomes significant in disc evolution (the artifact at that separation in Figure~\ref{fig:form_time}). \\

\noindent - Alternatively the planet migrates into the photoevaporative profile's peak and is then halted in its migration as photoevaporation, significantly aided interior(exterior) to the planet's orbit by the low(high) accretion efficiency of a more(less) massive planet, extends the leading(trailing) portion of the circumplanetary gap sufficiently to prevent resupply of the innermost disc. The innermost disc then drains onto the star, and photoevaporation disperses the remaining disc exterior to the newly formed hole.  
This process is discussed in more detail in Section~\ref{sec:migra_photoevap} and is the most common for planets that are halted in their migration (do not breach our 0.15 AU cutoff) in the X-ray and FUV cases of Figure~\ref{fig:form_time}, underscoring how the planet can facilitate conditions for photoevaporative disc dispersal in our model. The less common instances of this behavior under an EUV profile are predominantly lower mass giants whose rate of mass transfer across the circumplanetary gap is high enough to keep the inner disc resupplied when the circumplanetary gap first approaches the profile's peak region, but not enough to prevent dispersal shortly thereafter. \\

\noindent - Most common in the EUV and less frequent in the X-ray, the planet more quickly facilitates disc dispersal as its circumplanetary gap is just approaching the photoevaporative mass loss profile's peak rather than after the body has migrated interior to this peak. This stalls the object at separations exterior to that peak. \\

In the latter two scenarios, \textit{planets that are able to initiate photoevaporative disc dispersal in our model shorten the disc lifetime by order 0.1 -- 1 Myr} relative to the equivalent photoevaporating disc without a planet.

\subsection{Largest model uncertainties}
\label{sec:unc}
\noindent - The form of planetary accretion efficiency $\epsilon(M_{\rm p})$ is not well known and has been shown by AP12 to substantially affect the interaction between migrating planets and photoevaporative gap opening in the disc. They demonstrate that a constant, i.e., irrespective of planet mass, low $\epsilon/\epsilon_{\rm max} = 0.3$ fosters photoevaporative gap opening by strongly hindering gas flow across the circumplanetary gap and thus limiting mass transfer interior to the planet's orbit. This has a large effect on the distribution of final planet separations, yielding deserts and pile-ups of planets interior to 5 AU, and further shows stratification across planet mass (see AP12 Figure 2 and our Section~\ref{sec:migra_photoevap}). Conversely a large $\epsilon/\epsilon_{\rm max} = 1$ shows nominal influence of the planet on the time of photoevaporative gap opening. 
\textit{The disc lifetime in our models is therefore strongly sensitive to our prescription for $\epsilon(M_{\rm p})$} in Equation~\ref{eq:epsilon}.\\

\noindent - \cite{Wise18} use 2D simulations to in part test against results in AP12 and ER15, finding that the planetary migration timescale is significantly shorter than the timescale over which photoevaporation disperses the disc. This leads to their conclusion that photoevaporation has a negligible effect on migration and thus planet separations at the time of disc dispersal. Their comparison model with the X-ray photoevaporative profile in ER15 (the left panel of their Figure 8) shows that the addition of photoevaporation to a viscously evolving disc changes migrating giants' semimajor axes by $\leq 3\%$ over 200 kyr of evolution. 
By contrast our results suggest photoevaporation substantially affects planetary migration rates. In the example conditions of Figure~\ref{fig:discs}(g) -- (i), a 0.50, 2.75 or 5.00 {\mj} planet formed at 1 Myr at 5 AU in a disc without photoevaporation migrates to our 0.15 AU cutoff within 750 kyr. Under a $10^{-8}\ {\rm M_\odot\ yr^{-1}}$ X-ray or FUV photoevaporative profile the planet also ultimately reaches this cutoff, however the migration tracks show a 10\% departure from the case without photoevaporation after $260 - 530$ kyr (dependent on the planet mass and whether in the X-ray or FUV regime). Additionally the EUV profile in these panels halts the planet in its migration.
Generalizing these findings to the full range of the parameter space we explore, Figure~\ref{fig:form_time} shows that when the photoevaporative mass loss rate is sufficiently high ($\gtrsim 10 ^{-9}\ {\rm M_\odot\ yr^{-1}}$ under an X-ray or EUV profile; $\gtrsim 10 ^{-8}\ {\rm M_\odot\ yr^{-1}}$ in the FUV regime) and/or the planet is formed sufficiently late in the disc lifetime (dependent also on the body's mass), photoevaporation is able to stall the object in its migration external to our 0.15 AU cutoff.

We find that two differences between \cite{Wise18} and this work lead to the divergent results. First their 2D models that compare to ER15 are evolved for 200 kyr, while our simulations suggest that longer integrations are needed to observe the effect photoevaporation has on migration. Using as an example the cases in Figure~\ref{fig:discs}(g) -- (i), a $10^{-8}\ {\rm M_\odot\ yr^{-1}}$ X-ray photoevaporative profile requires $420$ kyr to cause a 10\% departure from the migration tracks in the equivalent simulations without photoevaporation. And although migration is a relatively fast process, we further find that in accordance with the two timescale behavior of disc dispersal \citep{Clarke2001}, the moment at which photoevaporation first dominates viscous accretion to drive disc evolution precipitates the stalling of planets in their migration. In our full population syntheses the mean time for photoevaporation to induce disc dispersal and stall bodies in their migration exterior to our 0.15 AU cutoff is 880 kyr after planet formation in the FUV regime, 390 kyr in the EUV and 850 kyr in the X-ray. We therefore expect that if the simulations in \cite{Wise18} were run over longer timescales, photoevaporation's effect on migration would become pronounced. 

This discrepancy between the two works is compounded by a second key difference; we include a prescription for mass accretion onto the object that acts to hinder mass transfer across the circumplanetary gap, while \cite{Wise18} treat the planet as having reached a terminal mass and so do not include accretion onto the body during migration. Mass accretion onto the object causes a dearth of mass on one side of the circumplanetary gap (along the outer edge of the gap for high mass giants, the inner edge for low mass), and once the circumplanetary gap overlaps with the EUV or X-ray photoevaporative profile's peak, if the planet was formed late enough for the disc surface density to have been sufficiently depleted beforehand, the circumplanetary gap is widened by photoevaporation, the inner disc drains onto the star, and the object stalled as the extant outer disc is dispersed. Mass accretion onto the planet therefore plays a critical role in our simulations in allowing photoevaporation to widen the circumplanetary gap and affect the planet's migration. \cite{Wise18} state that they expect this to be a contributing factor to the discrepancy between their results and those in AP12 and ER15, and we find the same.

Thus sufficiently long integration times are needed to capture photoevaporation's effect on migrating giants, and as discussed by AP12 and noted above, differing prescriptions for planetary accretion efficiency can contribute to divergent conclusions on the efficacy of photoevaporation to influence migration. This underscores the significant effects large uncertainties in disc model and planet migration parameters can have on simulation results from similar setups.\\

\noindent - The accuracy of Type II migration has recently been questioned by 2D hydrodynamic simulations, which have found \citep[e.g.,][]{Duffell2014,Duermann2015,Duermann2017} that migration can be a factor of a few faster than the Type II rate we employ here\footnote{Our code does not $enforce$ migration at the Type II rate, rather computes it self-consistently from the shape of the surface density and torque profiles. In a viscous code however, the surface density will always readjust so as to drive migration at the Type II rate \citep{Lin1986}, provided that the local disc mass is greater than the planet mass. When this latter condition is not met the migration physics do become dependent on the local disc mass and thus on the prescribed photoevaporative mass loss rate.}. In addition these simulations found that the accretion rate of material onto the planet modifies the migration rate. Though a comprehensive study of Type II migration for different planet masses on Myr timescales (as we simulate here) is still not available, and so we neglect this effect. By increasing the migration rate to agree with recent findings, our results would of course show more planets closer to the star.\\

\noindent - One may expect the distribution of planet separations we obtain to be sensitive to our choice of 5 AU as the formation location for all planets, with the true range of formation locations not yet strongly constrained. AP12 do test a set of models inserting planets at 10 AU and find no statistical difference between these and models using 5 AU, and simply for this reason we do not vary this parameter in our population syntheses.\\

\noindent - The exoplanet initial mass function (IMF) may be expected to influence our obtained separations; we use a flat IMF, while the observed IMF declines with planet mass \citep{Marcy2005}. ER15 find flat and $1/M_{\rm p}$ distributions yield qualitatively similar results; we retain our artificial, uniform distribution in order to draw enough planets in each mass bin to investigate mass-dependent trends in the population syntheses. \\

\noindent - In addition to photoevaporation, winds driven by magnetohydrodynamic (MHD) turbulence \citep[e.g.,][]{Bai2013,Armitage2013} likely also play a role in disc dispersal, with the relative effect of the two processes unclear. If the observational peak in Figure~\ref{fig:hist}(a) is an artifact of disc dispersal, this may not be due solely (or even primarily) to photoevaporation. Moreover dynamic evolution must (re)shape some fraction of observed planetary systems, partially or potentially wholly erasing the orbital signature left at the time of dispersal \citep[e.g.,][]{Ford2001,Moeckel2012b}.

\section{Conclusions}
\label{sec:conc}
We used a 1D viscous evolution code to simulate gas giant migration in a viscously evolving protoplanetary gas disc losing mass in either an FUV, EUV or X-ray driven photoevaporative wind induced by the host star. 
We found the photoevaporative mass loss profile's peak location and width have a strong effect in determining where and when gas giants are stalled in their migration, yielding characteristic features for each of the three energetic regimes: a severe deficit of $\lesssim$ 2 {\mj} planets interior to 5 AU in the FUV scenario, a sharp concentration of $\lesssim 3$ {\mj} planets between $\approx 1.5 - 2$ AU in the EUV case, and a relative excess of $\approx 2 - 3.5$ {\mj} giants interior to 0.5 AU in the X-ray model.
These features are not present in the sample of giants with minimum masses constrained by radial velocity measurements, and overall our simulated distributions fail to match observations. Our results are sensitive to a number of poorly constrained model parameters and are thus not meant for robust quantitative comparison to observations, but as an indication of trends.  

As future observations and theory place tighter constraints on models of the type used here, characteristic features of different photoevaporative regimes may serve as diagnostics to identify whether photoevaporative disc dispersal is driven primarily by FUV, EUV or X-ray irradiation. This may in turn elucidate the relative effect of photoevaporation in disc dispersal and aid an understanding of the extent to which disc processes determine the configuration of planetary systems.

\section*{Acknowledgements}
We thank U. Gorti for sharing her FUV models, as well as I. Pascucci and R. Alexander for the useful and constructive discussions. JJ thanks D. Henderson for his remarks. This research has made use of the Exoplanet Orbit Database and the Exoplanet Data Explorer at exoplanets.org. BE, JJ and GR acknowledge the support of the Munich Institute for Astro- and Particle Physics (MIAPP) of the DFG Cluster of Excellence \lq{}Origin and Structure of the Universe.\rq{} BE is supported by the Deutsche Forschungsgemeinschaft - Ref no. DFG ER 685/8-1 FOR 2634/1. This work has been supported by the DISCSIM project, grant agreement 341137 funded by the European Research Council under ERC-2013-ADG.

\bibliographystyle{mnras}
\bibliography{references.bib}

\begin{thebibliography}{}
\makeatletter
\relax
\def\mn@urlcharsother{\let\do\@makeother \do\$\do\&\do\#\do\^\do\_\do\%\do\~}
\def\mn@doi{\begingroup\mn@urlcharsother \@ifnextchar [ {\mn@doi@}
  {\mn@doi@[]}}
\def\mn@doi@[#1]#2{\def\@tempa{#1}\ifx\@tempa\@empty \href
  {http://dx.doi.org/#2} {doi:#2}\else \href {http://dx.doi.org/#2} {#1}\fi
  \endgroup}
\def\mn@eprint#1#2{\mn@eprint@#1:#2::\@nil}
\def\mn@eprint@arXiv#1{\href {http://arxiv.org/abs/#1} {{\tt arXiv:#1}}}
\def\mn@eprint@dblp#1{\href {http://dblp.uni-trier.de/rec/bibtex/#1.xml}
  {dblp:#1}}
\def\mn@eprint@#1:#2:#3:#4\@nil{\def\@tempa {#1}\def\@tempb {#2}\def\@tempc
  {#3}\ifx \@tempc \@empty \let \@tempc \@tempb \let \@tempb \@tempa \fi \ifx
  \@tempb \@empty \def\@tempb {arXiv}\fi \@ifundefined
  {mn@eprint@\@tempb}{\@tempb:\@tempc}{\expandafter \expandafter \csname
  mn@eprint@\@tempb\endcsname \expandafter{\@tempc}}}

\bibitem[\protect\citeauthoryear{Alexander \& Armitage}{Alexander \&
  Armitage}{2007}]{Alexander2007}
Alexander R.~D.,  Armitage P.~J.,  2007, \mn@doi [\mnras]
  {10.1111/j.1365-2966.2006.11341.x}, 375, 500

\bibitem[\protect\citeauthoryear{Alexander \& Armitage}{Alexander \&
  Armitage}{2009}]{Alexander2009}
Alexander R.~D.,  Armitage P.~J.,  2009, \mn@doi [\apj]
  {10.1088/0004-637X/704/2/989}, 704, 989

\bibitem[\protect\citeauthoryear{Alexander \& Pascucci}{Alexander \&
  Pascucci}{2012}]{Alexander2012a}
Alexander R.~D.,  Pascucci I.,  2012, \mn@doi [\mnras]
  {10.1111/j.1745-3933.2012.01243.x}, 422, L82

\bibitem[\protect\citeauthoryear{{Alexander}, {Clarke}  \&
  {Pringle}}{{Alexander} et~al.}{2006}]{Alexander2006b}
{Alexander} R.~D.,  {Clarke} C.~J.,   {Pringle} J.~E.,  2006, \mn@doi [\mnras]
  {10.1111/j.1365-2966.2006.10293.x}, \href
  {http://adsabs.harvard.edu/abs/2006MNRAS.369..216A} {369, 216}

\bibitem[\protect\citeauthoryear{{Alexander}, {Pascucci}, {Andrews}, {Armitage}
   \& {Cieza}}{{Alexander} et~al.}{2014}]{Alexander2014}
{Alexander} R.,  {Pascucci} I.,  {Andrews} S.,  {Armitage} P.,   {Cieza} L.,
  2014, \mn@doi [Protostars and Planets VI]
  {10.2458/azu_uapress_9780816531240-ch021}, \href
  {http://adsabs.harvard.edu/abs/2014prpl.conf..475A} {pp 475--496}

\bibitem[\protect\citeauthoryear{{Antonini}, {Hamers}  \&
  {Lithwick}}{{Antonini} et~al.}{2016}]{Antonini2016}
{Antonini} F.,  {Hamers} A.~S.,   {Lithwick} Y.,  2016, \mn@doi [\aj]
  {10.3847/0004-6256/152/6/174}, \href
  {http://adsabs.harvard.edu/abs/2016AJ....152..174A} {152, 174}

\bibitem[\protect\citeauthoryear{{Armitage}, {Livio}, {Lubow}  \&
  {Pringle}}{{Armitage} et~al.}{2002}]{Armitage2002}
{Armitage} P.~J.,  {Livio} M.,  {Lubow} S.~H.,   {Pringle} J.~E.,  2002,
  \mn@doi [\mnras] {10.1046/j.1365-8711.2002.05531.x}, \href
  {http://adsabs.harvard.edu/abs/2002MNRAS.334..248A} {334, 248}

\bibitem[\protect\citeauthoryear{{Armitage}, {Simon}  \& {Martin}}{{Armitage}
  et~al.}{2013}]{Armitage2013}
{Armitage} P.~J.,  {Simon} J.~B.,   {Martin} R.~G.,  2013, \mn@doi [\apjl]
  {10.1088/2041-8205/778/1/L14}, \href
  {http://adsabs.harvard.edu/abs/2013ApJ...778L..14A} {778, L14}

\bibitem[\protect\citeauthoryear{{Bai} \& {Stone}}{{Bai} \&
  {Stone}}{2013}]{Bai2013}
{Bai} X.-N.,  {Stone} J.~M.,  2013, \mn@doi [\apj]
  {10.1088/0004-637X/769/1/76}, \href
  {http://adsabs.harvard.edu/abs/2013ApJ...769...76B} {769, 76}

\bibitem[\protect\citeauthoryear{Bitsch, Lambrechts  \& Johansen}{Bitsch
  et~al.}{2015}]{Bitsch2015}
Bitsch B.,  Lambrechts M.,   Johansen A.,  2015, \aap, 582, A112

\bibitem[\protect\citeauthoryear{{Clarke}, {Gendrin}  \& {Sotomayor}}{{Clarke}
  et~al.}{2001}]{Clarke2001}
{Clarke} C.~J.,  {Gendrin} A.,   {Sotomayor} M.,  2001, \mn@doi [\mnras]
  {10.1046/j.1365-8711.2001.04891.x}, \href
  {http://adsabs.harvard.edu/abs/2001MNRAS.328..485C} {328, 485}

\bibitem[\protect\citeauthoryear{{Cumming}, {Butler}, {Marcy}, {Vogt}, {Wright}
   \& {Fischer}}{{Cumming} et~al.}{2008}]{Cumming2008}
{Cumming} A.,  {Butler} R.~P.,  {Marcy} G.~W.,  {Vogt} S.~S.,  {Wright} J.~T.,
   {Fischer} D.~A.,  2008, \mn@doi [\pasp] {10.1086/588487}, \href
  {http://adsabs.harvard.edu/abs/2008PASP..120..531C} {120, 531}

\bibitem[\protect\citeauthoryear{Davies, Adams, Armitage, Chambers, Ford,
  Morbidelli, Raymond  \& Veras}{Davies et~al.}{2013}]{Davies2013}
Davies M.~B.,  Adams F.~C.,  Armitage P.~J.,  Chambers J.~E.,  Ford E.~B.,
  Morbidelli A.,  Raymond S.~N.,   Veras D.,  2013, in , Protostars and Planets
  VI.
p.~23 (\mn@eprint {arXiv} {arXiv:1311.6816v1})

\bibitem[\protect\citeauthoryear{{Dawson} \& {Johnson}}{{Dawson} \&
  {Johnson}}{2018}]{Dawson2018}
{Dawson} R.~I.,  {Johnson} J.~A.,  2018, preprint, \href
  {http://adsabs.harvard.edu/abs/2018arXiv180106117D} {} (\mn@eprint {arXiv}
  {1801.06117})

\bibitem[\protect\citeauthoryear{{Dawson} \& {Murray-Clay}}{{Dawson} \&
  {Murray-Clay}}{2013}]{Dawson2013}
{Dawson} R.~I.,  {Murray-Clay} R.~A.,  2013, \mn@doi [\apjl]
  {10.1088/2041-8205/767/2/L24}, \href
  {http://adsabs.harvard.edu/abs/2013ApJ...767L..24D} {767, L24}

\bibitem[\protect\citeauthoryear{Duffell, Haiman, MacFadyen, D'Orazio  \&
  Farris}{Duffell et~al.}{2014}]{Duffell2014}
Duffell P.~C.,  Haiman Z.,  MacFadyen A.~I.,  D'Orazio D.~J.,   Farris B.~D.,
  2014, \mn@doi [\apj] {10.1088/2041-8205/792/1/L10}, 792, L10

\bibitem[\protect\citeauthoryear{D{\"{u}}rmann \& Kley}{D{\"{u}}rmann \&
  Kley}{2015}]{Duermann2015}
D{\"{u}}rmann C.,  Kley W.,  2015, \aap, 574, A52

\bibitem[\protect\citeauthoryear{{D{\"u}rmann} \& {Kley}}{{D{\"u}rmann} \&
  {Kley}}{2017}]{Duermann2017}
{D{\"u}rmann} C.,  {Kley} W.,  2017, \mn@doi [\aap]
  {10.1051/0004-6361/201629074}, \href
  {http://adsabs.harvard.edu/abs/2017A%26A...598A..80D} {598, A80}

\bibitem[\protect\citeauthoryear{{Ercolano} \& {Pascucci}}{{Ercolano} \&
  {Pascucci}}{2017}]{ErcolanoPascucci2017}
{Ercolano} B.,  {Pascucci} I.,  2017, \mn@doi [Royal Society Open Science]
  {10.1098/rsos.170114}, \href
  {http://adsabs.harvard.edu/abs/2017RSOS....470114E} {4, 170114}

\bibitem[\protect\citeauthoryear{Ercolano \& Rosotti}{Ercolano \&
  Rosotti}{2015}]{Ercolano2015}
Ercolano B.,  Rosotti G.~P.,  2015, \mn@doi [\mnras] {10.1093/mnras/stv833},
  450, 3008

\bibitem[\protect\citeauthoryear{Ercolano, Clarke  \& Drake}{Ercolano
  et~al.}{2009}]{Ercolano2009}
Ercolano B.,  Clarke C.~J.,   Drake J.~J.,  2009, \mn@doi [\apj]
  {10.1088/0004-637X/699/2/1639}, 699, 1639

\bibitem[\protect\citeauthoryear{Fedele, van~den Ancker, Henning, Jayawardhana
  \& Oliveira}{Fedele et~al.}{2010}]{Fedele2010}
Fedele D.,  van~den Ancker M.~E.,  Henning T.,  Jayawardhana R.,   Oliveira
  J.~M.,  2010, \mn@doi [\aap] {10.1051/0004-6361/200912810}, 510, 72

\bibitem[\protect\citeauthoryear{{Font}, {McCarthy}, {Johnstone}  \&
  {Ballantyne}}{{Font} et~al.}{2004}]{Font2004}
{Font} A.~S.,  {McCarthy} I.~G.,  {Johnstone} D.,   {Ballantyne} D.~R.,  2004,
  \mn@doi [\apj] {10.1086/383518}, \href
  {http://adsabs.harvard.edu/abs/2004ApJ...607..890F} {607, 890}

\bibitem[\protect\citeauthoryear{{Ford}, {Havlickova}  \& {Rasio}}{{Ford}
  et~al.}{2001}]{Ford2001}
{Ford} E.~B.,  {Havlickova} M.,   {Rasio} F.~A.,  2001, \mn@doi [\icarus]
  {10.1006/icar.2001.6588}, \href
  {http://adsabs.harvard.edu/abs/2001Icar..150..303F} {150, 303}

\bibitem[\protect\citeauthoryear{Goldreich \& Tremaine}{Goldreich \&
  Tremaine}{1980}]{Goldreich1980}
Goldreich P.,  Tremaine S.,  1980, \mn@doi [\apj] {10.1086/158356}, 241, 425

\bibitem[\protect\citeauthoryear{{Gorti} \& {Hollenbach}}{{Gorti} \&
  {Hollenbach}}{2009}]{Gorti2009a}
{Gorti} U.,  {Hollenbach} D.,  2009, \mn@doi [\apj]
  {10.1088/0004-637X/690/2/1539}, \href
  {http://adsabs.harvard.edu/abs/2009ApJ...690.1539G} {690, 1539}

\bibitem[\protect\citeauthoryear{{Gorti}, {Dullemond}  \& {Hollenbach}}{{Gorti}
  et~al.}{2009}]{Gorti2009}
{Gorti} U.,  {Dullemond} C.~P.,   {Hollenbach} D.,  2009, \mn@doi [\apj]
  {10.1088/0004-637X/705/2/1237}, \href
  {http://adsabs.harvard.edu/abs/2009ApJ...705.1237G} {705, 1237}

\bibitem[\protect\citeauthoryear{{G{\"u}del} et~al.,}{{G{\"u}del}
  et~al.}{2007}]{Gudel2007}
{G{\"u}del} M.,  et~al., 2007, \mn@doi [\aap] {10.1051/0004-6361:20065724},
  \href {http://adsabs.harvard.edu/abs/2007A%26A...468..353G} {468, 353}

\bibitem[\protect\citeauthoryear{{Han}, {Wang}, {Wright}, {Feng}, {Zhao},
  {Fakhouri}, {Brown}  \& {Hancock}}{{Han} et~al.}{2014}]{Han2014}
{Han} E.,  {Wang} S.~X.,  {Wright} J.~T.,  {Feng} Y.~K.,  {Zhao} M.,
  {Fakhouri} O.,  {Brown} J.~I.,   {Hancock} C.,  2014, \mn@doi [\pasp]
  {10.1086/678447}, \href {http://adsabs.harvard.edu/abs/2014PASP..126..827H}
  {126, 827}

\bibitem[\protect\citeauthoryear{{Hartmann}, {Calvet}, {Gullbring}  \&
  {D'Alessio}}{{Hartmann} et~al.}{1998}]{Hartmann1998}
{Hartmann} L.,  {Calvet} N.,  {Gullbring} E.,   {D'Alessio} P.,  1998, \mn@doi
  [\apj] {10.1086/305277}, \href
  {http://adsabs.harvard.edu/abs/1998ApJ...495..385H} {495, 385}

\bibitem[\protect\citeauthoryear{Ida \& Lin}{Ida \& Lin}{2004}]{Ida2004a}
Ida S.,  Lin D. N.~C.,  2004, \mn@doi [\apj] {10.1086/381724}, 604, 388

\bibitem[\protect\citeauthoryear{Kley \& Nelson}{Kley \&
  Nelson}{2012}]{Kley2012a}
Kley W.,  Nelson R.~P.,  2012, \mn@doi [\araa]
  {10.1146/annurev-astro-081811-125523}, 50, 211

\bibitem[\protect\citeauthoryear{{Laughlin} \& {Adams}}{{Laughlin} \&
  {Adams}}{1997}]{Laughlin1997}
{Laughlin} G.,  {Adams} F.~C.,  1997, \mn@doi [\apjl] {10.1086/311056}, \href
  {http://adsabs.harvard.edu/abs/1997ApJ...491L..51L} {491, L51}

\bibitem[\protect\citeauthoryear{Lin \& Papaloizou}{Lin \&
  Papaloizou}{1986}]{Lin1986}
Lin D. N.~C.,  Papaloizou J. C.~B.,  1986, \apj, 309, 846

\bibitem[\protect\citeauthoryear{Lynden-Bell \& Pringle}{Lynden-Bell \&
  Pringle}{1974}]{Lynden-Bell1974}
Lynden-Bell D.,  Pringle J.~E.,  1974, \mnras, 168, 603

\bibitem[\protect\citeauthoryear{{Marcy}, {Butler}, {Fischer}, {Vogt},
  {Wright}, {Tinney}  \& {Jones}}{{Marcy} et~al.}{2005}]{Marcy2005}
{Marcy} G.,  {Butler} R.~P.,  {Fischer} D.,  {Vogt} S.,  {Wright} J.~T.,
  {Tinney} C.~G.,   {Jones} H.~R.~A.,  2005, \mn@doi [Progress of Theoretical
  Physics Supplement] {10.1143/PTPS.158.24}, \href
  {http://adsabs.harvard.edu/abs/2005PThPS.158...24M} {158, 24}

\bibitem[\protect\citeauthoryear{{Moeckel} \& {Armitage}}{{Moeckel} \&
  {Armitage}}{2012}]{Moeckel2012b}
{Moeckel} N.,  {Armitage} P.~J.,  2012, \mn@doi [\mnras]
  {10.1111/j.1365-2966.2011.19699.x}, \href
  {http://adsabs.harvard.edu/abs/2012MNRAS.419..366M} {419, 366}

\bibitem[\protect\citeauthoryear{Mordasini, Molli{\`{e}}re, Dittkrist, Jin  \&
  Alibert}{Mordasini et~al.}{2015}]{Mordasini2015}
Mordasini C.,  Molli{\`{e}}re P.,  Dittkrist K.-M.,  Jin S.,   Alibert Y.,
  2015, \mn@doi [International Journal of Astrobiology]
  {10.1017/S1473550414000263}, 14, 201

\bibitem[\protect\citeauthoryear{Owen, Ercolano, Clarke  \& Alexander}{Owen
  et~al.}{2010}]{Owen2010}
Owen J.~E.,  Ercolano B.,  Clarke C.~J.,   Alexander R.~D.,  2010, \mn@doi
  [\mnras] {10.1111/j.1365-2966.2009.15771.x}, 401, 1415

\bibitem[\protect\citeauthoryear{Owen, Ercolano  \& Clarke}{Owen
  et~al.}{2011}]{Owen2011}
Owen J.~E.,  Ercolano B.,   Clarke C.~J.,  2011, \mn@doi [\mnras]
  {10.1111/j.1365-2966.2010.17818.x}, 412, 13

\bibitem[\protect\citeauthoryear{Owen, Clarke  \& Ercolano}{Owen
  et~al.}{2012}]{Owen2012}
Owen J.~E.,  Clarke C.~J.,   Ercolano B.,  2012, \mn@doi [\mnras]
  {10.1111/j.1365-2966.2011.20337.x}, 422, 1880

\bibitem[\protect\citeauthoryear{{Pascucci} \& {Sterzik}}{{Pascucci} \&
  {Sterzik}}{2009}]{Pascucci2009}
{Pascucci} I.,  {Sterzik} M.,  2009, \mn@doi [\apj]
  {10.1088/0004-637X/702/1/724}, \href
  {http://adsabs.harvard.edu/abs/2009ApJ...702..724P} {702, 724}

\bibitem[\protect\citeauthoryear{{Pascucci}, {Ricci}, {Gorti}, {Hollenbach},
  {Hendler}, {Brooks}  \& {Contreras}}{{Pascucci} et~al.}{2014}]{Pascucci2014}
{Pascucci} I.,  {Ricci} L.,  {Gorti} U.,  {Hollenbach} D.,  {Hendler} N.~P.,
  {Brooks} K.~J.,   {Contreras} Y.,  2014, \mn@doi [\apj]
  {10.1088/0004-637X/795/1/1}, \href
  {http://adsabs.harvard.edu/abs/2014ApJ...795....1P} {795, 1}

\bibitem[\protect\citeauthoryear{Petrovich}{Petrovich}{2015}]{Petrovich2015a}
Petrovich C.,  2015, \mn@doi [\apj] {10.1088/0004-637X/805/1/75}, 805, 75

\bibitem[\protect\citeauthoryear{Preibisch et~al.,}{Preibisch
  et~al.}{2005}]{Preibisch2005}
Preibisch T.,  et~al., 2005, \mn@doi [\apjs] {10.1086/432891}, 160, 401

\bibitem[\protect\citeauthoryear{{Pringle}, {Verbunt}  \& {Wade}}{{Pringle}
  et~al.}{1986}]{Pringle1986}
{Pringle} J.~E.,  {Verbunt} F.,   {Wade} R.~A.,  1986, \mn@doi [\mnras]
  {10.1093/mnras/221.2.169}, \href
  {http://adsabs.harvard.edu/abs/1986MNRAS.221..169P} {221, 169}

\bibitem[\protect\citeauthoryear{Rosotti, Ercolano, Owen  \& Armitage}{Rosotti
  et~al.}{2013}]{Rosotti2013}
Rosotti G.~P.,  Ercolano B.,  Owen J.~E.,   Armitage P.~J.,  2013, \mn@doi
  [\mnras] {10.1093/mnras/sts725}, 430, 1392

\bibitem[\protect\citeauthoryear{{Ruden}}{{Ruden}}{2004}]{Ruden2004}
{Ruden} S.~P.,  2004, \mn@doi [\apj] {10.1086/382524}, \href
  {http://adsabs.harvard.edu/abs/2004ApJ...605..880R} {605, 880}

\bibitem[\protect\citeauthoryear{{Schlaufman} \& {Winn}}{{Schlaufman} \&
  {Winn}}{2016}]{Schlaufman2016}
{Schlaufman} K.~C.,  {Winn} J.~N.,  2016, \mn@doi [\apj]
  {10.3847/0004-637X/825/1/62}, \href
  {http://adsabs.harvard.edu/abs/2016ApJ...825...62S} {825, 62}

\bibitem[\protect\citeauthoryear{Shakura \& Sunyaev}{Shakura \&
  Sunyaev}{1973}]{Shakura1973}
Shakura N.~I.,  Sunyaev R.~A.,  1973, \aap, 24, 337

\bibitem[\protect\citeauthoryear{{Sicilia-Aguilar}, {Henning}  \&
  {Hartmann}}{{Sicilia-Aguilar} et~al.}{2010}]{Sicilia-Aguilar2010}
{Sicilia-Aguilar} A.,  {Henning} T.,   {Hartmann} L.~W.,  2010, \mn@doi [\apj]
  {10.1088/0004-637X/710/1/597}, \href
  {http://adsabs.harvard.edu/abs/2010ApJ...710..597S} {710, 597}

\bibitem[\protect\citeauthoryear{{Veras} \& {Armitage}}{{Veras} \&
  {Armitage}}{2004}]{Veras2004}
{Veras} D.,  {Armitage} P.~J.,  2004, \mn@doi [\mnras]
  {10.1111/j.1365-2966.2004.07239.x}, \href
  {http://adsabs.harvard.edu/abs/2004MNRAS.347..613V} {347, 613}

\bibitem[\protect\citeauthoryear{{Winn}}{{Winn}}{2018}]{Winn2018}
{Winn} J.~N.,  2018, preprint, \href
  {http://adsabs.harvard.edu/abs/2018arXiv180108543W} {} (\mn@eprint {arXiv}
  {1801.08543})

\bibitem[\protect\citeauthoryear{{Wise} \& {Dodson-Robinson}}{{Wise} \&
  {Dodson-Robinson}}{2018}]{Wise18}
{Wise} A.~W.,  {Dodson-Robinson} S.~E.,  2018, \mn@doi [\apj]
  {10.3847/1538-4357/aaaae5}, \href
  {http://adsabs.harvard.edu/abs/2018ApJ...855..145W} {855, 145}

\bibitem[\protect\citeauthoryear{van Leer}{van Leer}{1977}]{vanLeer77}
van Leer B.,  1977, \mn@doi [Journal of Computational Physics]
  {10.1016/0021-9991(77)90095-X}, 23, 276

\makeatother
\end{thebibliography}

\bsp
\label{lastpage}
\end{document}